\documentstyle[aaspp4]{article}
\begin{document}
%
%%%%%%%%%%%%%%%%%%%%%%%%%%%%%%%%%%%%%%%%%%%%%%%%%%%%%%%%%%%%%%%%%
%%%%%% (1) Title Page   %%%%%%%%%%%%%%%%%%%%%%%%%%%%%%%%%%%%%%%%%%
%%%%%%%%%%%%%%%%%%%%%%%%%%%%%%%%%%%%%%%%%%%%%%%%%%%%%%%%%%%%%%%%%%

\title{Photodissociative Regulation of Star Formation in Metal-Free
  Pregalactic Clouds}  
\author{Kazuyuki Omukai and Ryoichi Nishi}
\affil{Department of Physics, Kyoto University, Kyoto 606-8502, Japan}

%%%%%%%%%%%%%%%%%%%%%%%%%%%%%%%%%%%%%%%%%%%%%%%%%%%%%%%%%%%%%%%%%%
%%%%%% (2) Abstract & Subject Heading  %%%%%%%%%%%%%%%%%%%%%%%%%%%
%%%%%%%%%%%%%%%%%%%%%%%%%%%%%%%%%%%%%%%%%%%%%%%%%%%%%%%%%%%%%%%%%%

\begin{abstract}
We study the H$_{2}$ photodissociation regions around OB
stars in primordial gas clouds whose virial temperatures are between a
few hundred and a few thousand Kelvin.
In such small objects, a single O star can photodissociate a mass
equal to that of the cloud itself.
As a result, the clouds deplete their molecular coolant and cannot cool in a
free-fall time, and subsequent star formation is totally quenched.
This indicates that stars do not form efficiently in small objects and
that these objects contribute little to the reionization of the universe. 
\end{abstract}


%%%%%%%%%%%%%%%%%%%%%%%%%%%%%%%%%%%%%%%%%%%%%%%%%%%%%%%%%%%%%%%%%%
%%%%%% (4) Text & Ackowledgment %%%%%% %%%%%%%%%%%%%%%%%%%%%%%%%%%
%%%%%%%%%%%%%%%%%%%%%%%%%%%%%%%%%%%%%%%%%%%%%%%%%%%%%%%%%%%%%%%%%%
\newpage

\section{Introduction}
After standard recombination at $z \simeq 1100$, the universe is
considered to be reionized at some redshift $z \gtrsim 5$ from the
negative results of Gunn-Peterson experiments in quasar spectra.
It has been pointed out that first stars can play an important role in
the reionization of the universe (e.g., Couchman \& Rees 1986).  
This scenario has been investigated in detail using numerical
simulations (e.g., Gnedin \& Ostriker 1997) 
or semi-analytical models (e.g., Fukugita \& Kawasaki 1994; Haiman \&
Loeb 1997).  
In the latter models, it is crucial to know whether a cloud of mass $M$
that virializes at redshift $z$ can cool or not, i.e., if star
formation occurs or not. 
     
Haiman, Thoul, \& Loeb (1996) investigated this problem (see also
Tegmark et al. 1997).  
They found that molecular cooling plays a crucial role for clouds with $T_{\rm
  vir} \lesssim 10^{4} {\rm K}$ (``small'' pregalactic clouds, hereafter) at
$10 \lesssim z_{\rm vir} \lesssim 100$, where $T_{\rm vir}$ and $z_{\rm
  vir}$ are the virial temperature and the redshift at virialization
respectively, and determined the minimum mass of the virialized cloud
must have in order to cool in a Hubble time.  

On the other hand, Haiman, Rees, \& Loeb (1997) pointed out that in the
presence of ultraviolet (UV) background radiation at the level needed to
ionize the universe, molecular hydrogen is photodissociated by far ultraviolet
(FUV) photons, whose radiation energy is less than the Lyman limit, in
small pregalactic clouds.   
Thus, they asserted that molecular hydrogen in small pregalactic clouds 
is universally destroyed long before the reionization of the universe. 
In their reionization model, Haiman \& Loeb (1997) assumed that only
objects with virial temperature above $10^{4}$ K can cool in a Hubble
time, owing to atomic cooling and star formation occurs subsequently.

Recently, Ciardi, Ferrara, \& Abel (1999) found that the
photodissociated regions are not large enough to overlap at $z \simeq
20-30$.
In the same redshift range, the flux of FUV background is well below the
threshold required by Haiman et al.(1997) to prevent the collapse of the
clouds.   
  
However, molecular hydrogen in a virialized cloud is photodissociated
not only by external FUV background radiation, but also by FUV photons
produced by massive stars within the cloud.
Here, we assess the negative feedback of massive star formation on
molecular hydrogen formation in a primordial cloud. 

\section{Region of Influence of an OB star}
Around an OB star, hydrogen is photoionized, and an HII region is formed.  
Ionizing photons hardly escape from the HII region, but photons whose
radiation energy are below the Lyman limit can get away.
Such FUV photons photodissociate molecular hydrogen, and a
photodissociation region (PDR) is formed just outside the HII region.
In this section, we study how much mass in a primordial cloud is
affected by such FUV photons from an OB star and, as a result, becomes
unable to cool in a free-fall time owing to the lack of the coolant.   
 
We consider a small pregalactic cloud of primordial composition. 
In such an object, H$_{2}$ is formed mainly by the H$^{-}$ process at
$z \lesssim 100$: 
\begin{eqnarray}
\label{eq:Hm1}
{\rm H}+e^{-} &\rightarrow & {\rm H^{-}}+\gamma ; \\
\label{eq:Hm2}
{\rm H}+{\rm H^{-}} &\rightarrow & {\rm H_{2}}+e^{-}.
\end{eqnarray}
At $z \gtrsim 100$, H$^{-}$ is predominantly photodissociated by CMB photons
before the reaction (\ref{eq:Hm2}) proceeds.
On the other hand, in a PDR, photodissociation of H$^{-}$ by UV
radiation from an OB star does not dominate the reaction (\ref{eq:Hm2})
except in the vicinity of the star.
Then we neglect the photodissociation of H$^{-}$.

The rate-determining stage of the H$^{-}$ process is the reaction
 (\ref{eq:Hm1}), whose rate coefficient $k_{\rm H^{-}}$  is (de Jong 1972)
\begin{equation}
  k_{\rm H^{-}}=1.0 \times 10^{-18} T~{\rm s^{-1} cm^{3}}. 
\end{equation}
In a PDR, H$_{2}$ is dissociated mainly via the two-step photodissociation
process: 
\begin{equation}
  {\rm H_2}+\gamma  \rightarrow {\rm H_{2}^*} \rightarrow 2{\rm H},
\end{equation}
whose rate coefficient $k_{\rm 2step}$ is given by 
(Kepner, Babul, \& Spergel 1997; Draine \& Bertoldi 1996)
\begin{equation}
  k_{\rm 2step}=1.13 \times 10^{8} F_{\rm LW}~{\rm s^{-1}}. 
\end{equation}
Here $F_{\rm LW}~({\rm ergs~s^{-1}cm^{-2}Hz^{-1}})$ is the averaged
radiation flux in the Lyman and Werner (LW) bands and can be written as 
\begin{equation}
F_{\rm LW}=F_{\rm LW,ex} f_{\rm shield},
\end{equation}
where $F_{\rm LW,ex}$ is the incident flux into the PDR at
12.4 eV and the shielding factor $f_{\rm shield}$ is given by
(Draine \& Bertoldi 1996)
\begin{equation}
\label{eq:fsh}
f_{\rm shield}={\rm min} \left[ 1,(\frac{N_{\rm H_2}}{10^{14}})^{-0.75}
\right].
\end{equation}

The timescale in which the H$_{2}$ fraction reaches the equilibrium
value is given by 
\begin{equation}
  t_{\rm dis}=k_{\rm 2step}^{-1}.
\end{equation}
In the presence of FUV radiation, if the temperature and density were
fixed, the H$_{2}$ fraction $f$ initially would increase and reach the
equilibrium value for a temporal ionization fraction after $\sim t_{\rm
  dis}$, and then it would decline as ionization fraction decreased as a 
result of recombination. 
Actually, if the pregalactic cloud can once produce a sufficient amount 
of molecular hydrogen to cool in a free-fall time, the cloud can
collapse and star formation occurs subsequently.
Note that because of their low ionization degree, inverse Compton
cooling by CMB photons is not effective in small objects with $T_{\rm
  vir} \lesssim 10^{4}$ K.  

We investigate here how much mass around an OB star is affected by the 
photodissociating FUV radiation from the star and becomes unable to cool in a
free-fall time.
In particular, we seek the lower bound of such mass.

The equilibrium number density of H$_{2}$ under ionization degree $x$ is
\begin{eqnarray}
\label{eq:neq}
n_{\rm H_2} &=& \frac{k_{\rm H^{-}}}{k_{\rm 2step}}x n^{2} \\
&=& 0.88 \times 10^{-26} x F_{\rm LW}^{-1} T  n^{2}.
\end{eqnarray}
Near the star, $F_{\rm LW}$ is so large that the dissociation time
$t_{\rm dis}$ is smaller than the recombination time $t_{\rm
  rec}=(k_{\rm rec} x_{\rm i} n)^{-1}$, where $k_{\rm rec}$ is the
recombination coefficient, $x_{\rm i}$ is the ionization degree at
virialization, and $n$ is the number density of hydrogen nuclei.
Then the chemical equilibrium between above processes is reached before
significant recombination proceeds.
Far distant from the star, the recombination proceeds before the
molecular fraction reaches the equilibrium value and the ionization
degree significantly diminishes. 
However, since we are seeking how much mass is at least affected by the 
photodissociating FUV radiation from the star, we use the equilibrium
value (\ref{eq:neq}) with initial ionization degree $x=x_{\rm i}$ 
as the H$_{2}$ number density. 
 
Consider an OB star, which radiates at the rate of  
$L_{\rm LW}$ [ergs s$^{-1}$ Hz$^{-1}$] in the LW bands.
For an O5 star, whose mass is 40 $M_{\sun}$, $L_{\rm LW} \simeq
10^{24}$ ergs s$^{-1}$ Hz$^{-1}$. 
At the point whose distance from the star is $r$, the averaged flux in
the Lyman and Werner bands is approximately given by 
\begin{equation}
F_{\rm LW}=\frac{L_{\rm LW}}{4 \pi r^{2}} f_{\rm shield}.
\end{equation}

Using above relations, we obtain the H$_2$ column density between the
star and the point whose distance from the star is $r$:
\begin{equation}
\label{eq:column}
N_{\rm H_2}= \left\{
\begin{array}{ll}
C V & \mbox{($N_{\rm H_2}<10^{14} {\rm cm^{-3}}$)} \\
10^{14}[0.25 (C V/10^{14}) + 0.75]^{4} &
\mbox{($N_{\rm H_2}>10^{14}{\rm cm^{-3}}$)}
\end{array}
\right.
\end{equation}
where
\begin{eqnarray}
C &=& 0.88 \times 10^{-26} x_{\rm i}L_{\rm LW}^{-1} T n^{2},\\
V &=& \frac{4 \pi}{3} r^{3}.
\end{eqnarray}
Here we have assumed $n=$const. in space, for simplicity.
We define here the region of influence around a star as that where the
cooling time $t_{\rm cool}=\frac{(3/2)kT}{fn \Lambda_{\rm H_2}}$ becomes
larger than the free-fall time $t_{\rm ff}=(\frac{3 \pi \Omega_{\rm b}}
{32G m_{\rm p} n})^{1/2}$ as a result of the photodissociation of
molecular hydrogen, where $f=n_{\rm H_{2}}/n$ is the H$_{2}$ concentration,
$\Lambda_{\rm H_2}$ is the cooling function of molecular hydrogen, and
$\Omega_{\rm b}$ is the baryon mass fraction.  

The condition $t_{\rm cool} > t_{\rm ff}$ is satisfied as long as the
H$_2$ fraction  
\begin{eqnarray}
  f < f^{\rm (cool)} &=& (\frac{24 G m_{\rm p}}{\pi})^{1/2}
  \frac{kT}{\Omega_{\rm b}^{1/2} n^{1/2} \Lambda_{\rm H_2}}\\
  &=& 1 \times 10^{-3} (\frac{n}{1 {\rm cm^{-3}}})^{-1/2} (\frac{T}{10^3
  {\rm K}})^{-3} (\frac{\Omega_{\rm b}}{0.05})^{-1/2},  
\end{eqnarray}
where we used our fit to the Martin, Schwarz,\& Mandy (1996) H$_{2}$
cooling function
\begin{equation}
\Lambda_{\rm H_2} \simeq 4 \times 10^{-25}
(\frac{T}{1000{\rm K}})^{4} {\rm ergs~s^{-1}cm^{3}}
\end{equation}
for the low temperature ($600{\rm K} \lesssim T \lesssim 3000{\rm K}$)
and low density ($n \lesssim 10^{4} {\rm cm^{-3}}$) regime.
The same condition leads to the condition on the averaged flux in the
LW bands with equation (\ref{eq:neq}):
\begin{eqnarray}
  F_{\rm LW} > F_{\rm LW}^{\rm (cool)}
&=& \frac{k_{\rm H^{-}} x_{\rm i}n }{1.13 \times 10^{8} f^{(\rm cool)}}\\
&=& 0.7 \times 10^{-24} {\rm ergs~s^{-1} cm^{-2} Hz^{-1}}
(\frac{x_{\rm i}}{10^{-4}}) (\frac{n}{1 {\rm 
cm^{-3}}})^{3/2} (\frac{T}{10^3 {\rm K}})^{4} (\frac{\Omega_{\rm
b}}{0.05})^{1/2}.
\end{eqnarray}

Corresponding to this critical LW flux $F_{\rm LW}^{\rm (cool)}$, a
critical radius $r^{\rm (cool)}$ is determined by the relation  
\begin{equation}
  F_{\rm LW}[r^{\rm (cool)}]=F_{\rm LW}^{\rm (cool)}.
\end{equation}
Actually, the timescale for molecular hydrogen to reach the equilibrium
value at $r=r^{\rm (cool)}$,  
\begin{equation}
  t_{\rm dis}[r^{\rm (cool)}]=4.0 \times 10^{8} {\rm yr} 
(\frac{x_{\rm i}}{10^{-4}})^{-1}
(\frac{n}{1 {\rm cm^{-3}}})^{-3/2} 
(\frac{T}{10^3 {\rm K}})^{-4} 
(\frac{\Omega_{\rm b}}{0.05})^{-1/2}, 
\end{equation}
is longer than the lifetime of an OB star $t_{\rm OB} \sim 3 \times
10^{6} $ yr. 
This means that as far region as $r=r^{\rm (cool)}$ is rarely affected
within the lifetime of a single massive star.
We define here another critical LW flux $F_{\rm LW}^{(\rm eq)}$
and its corresponding radius $r^{\rm (eq)}$ where the timescale for
molecular hydrogen to reach the equilibrium value $t_{\rm dis}$ becomes
equal to the lifetime of an OB star;  
\begin{equation}
F_{\rm LW}^{(\rm eq)}=0.93 \times 10^{-22} {\rm ergs~s^{-1} cm^{-2} Hz^{-1}}
(\frac{t_{\rm OB}}{3 \times 10^6 {\rm yr}})^{-1},
\end{equation}
\begin{equation}
F_{\rm LW}[r^{\rm (eq)}]=F_{\rm LW}^{\rm (eq)}.
\end{equation}

In the region of influence, we require that two conditions be met: (1) the
  cooling time for the equilibrium H$_{2}$ fraction is longer than
  the free-fall time (i.e., $t_{\rm cool}[f=f^{\rm (eq)}]>t_{\rm
  ff}$,where $f^{\rm (eq)}$ is the equilibrium H$_{2}$ fraction) 
   and (2) the equilibrium H$_{2}$ fraction is reached within the
  lifetime of the central star (i.e., $t_{\rm dis}< t_{\rm OB}$).
Hence the radius of influence $r^{\rm (inf)}$ is determined by the
  smaller one of either $r^{\rm (cool)}$ or $r^{\rm (eq)}$;
\begin{equation}
r^{\rm (inf)}={\rm min}[r^{\rm (cool)},r^{\rm (eq)}].
\end{equation}

Note the LW flux $F_{\rm LW}$ at which $t_{\rm dis}$ becomes equal to
$t_{\rm rec}$ is $2 \times 10^{-24} (x_{\rm i}/10^{-4})(n/1 {\rm
  cm^{-3}})(T/10^{3} {\rm K})^{-0.64} {\rm ergs~s^{-1} cm^{-2} Hz^{-1}}$. 
Here we used the recombination coefficient $k_{\rm rec}=1.88 \times
10^{-10} T^{-0.64}~{\rm s^{-1} cm^{3}}$ (Hutchins 1976).
Therefore, even at the edge of the region of influence, the condition
$t_{\rm dis}<t_{\rm rec}$ is usually satisfied.
In this case, the chemical equilibrium value for H$_2$ is reached before
the ionization degree significantly decreases from the initial value
$x_{\rm i}$. 

If the self-shielding of LW band photons could be
neglected, the radius $r^{\rm (cool)}$ would be given by
\begin{eqnarray}
\label{eq:rinf}
  r^{({\rm cool})}=r^{({\rm cool})}_{\rm no-sh}
  &=&[\frac{L_{\rm LW}}{4 \pi F_{\rm LW}^{\rm (cool)}}]^{1/2} \\
  &=&3.4 \times 10^{23} {\rm cm} (\frac{x_{\rm i}}{10^{-4}})^{-1/2}
  (\frac{L_{\rm LW}}{10^{24} {\rm ergs~s^{-1} Hz^{-1}}})^{1/2}
  (\frac{T}{10^3 {\rm K}})^{-2} (\frac{n}{1 {\rm cm^{-3}}})^{-3/4}
  (\frac{\Omega_{\rm b}}{0.05})^{-1/4}.
\end{eqnarray}
This expression is valid only when $N_{\rm H_{2}}<10^{14} {\rm
  cm^{-2}}$.

When the H$_2$ column density $N_{\rm H_2}$ becomes larger than
$10^{14} {\rm cm^{-2}}$, self-shielding of LW band photons 
begins as can be seen from equation (\ref{eq:fsh}).
Here, we define the shielding radius $r_{\rm sh}$ as the radius where $N_{\rm
  H_2}(r_{\rm sh})=10^{14} {\rm cm^{-2}}$.
Using equation (\ref{eq:column}), the shielding radius is 
\begin{eqnarray}
\label{eq:rsh}
  r_{\rm sh}&=&[\frac{10^{14}}{(4 \pi/3) C}]^{1/3} \\
  &=&3.0 \times 10^{21} {\rm cm} (\frac{x_{\rm i}}{10^{-4}})^{-1/3}
  (\frac{L_{\rm LW}}{10^{24} {\rm ergs~s^{-1} Hz^{-1}}})^{1/3}
  (\frac{T}{10^3 {\rm K}})^{-1/3} (\frac{n}{1 {\rm cm^{-3}}})^{-2/3}.
\end{eqnarray}
When the self-shielding becomes important, $N_{\rm H_2}$ increases and
  $F_{\rm LW}$ decreases rapidly with $r$.  
Then $r^{\rm (cool)}$ is not much larger than $r_{\rm sh}$.
In such a case, we put $r^{\rm (cool)}=r_{\rm sh}$ as a lower bound.
Then $r^{\rm (cool)}$ is given by the lesser one of those given by
  equations (\ref{eq:rinf}) or (\ref{eq:rsh}).

In all the same way as above, we can obtain the value of $r^{({\rm
    eq})}$;
\begin{equation}
  r^{({\rm eq})}={\rm min}[r^{({\rm eq})}_{\rm no-sh},r_{\rm sh}],
\end{equation}
where 
\begin{eqnarray}
\label{eq:req}
r^{({\rm eq})}_{\rm no-sh}
&=&[\frac{L_{\rm LW}}{4 \pi F_{\rm LW}^{\rm (eq)}}]^{1/2} \\
&=&2.9 \times 10^{22}
(\frac{L_{\rm LW}}{10^{24} {\rm ergs~s^{-1} Hz^{-1}}})^{1/2}
(\frac{t_{\rm OB}}{3 \times 10^{6} {\rm yr}})^{1/2}.
\end{eqnarray}

The baryonic mass within the region of influence $M_{\rm b}^{\rm (inf)}$
is then 
\begin{eqnarray}
M_{\rm b}^{\rm (inf)} &=& \frac{4 \pi}{3} n m_{\rm p} {r^{\rm (inf)}}^{3}\\
\label{eq:Minf}
 &=& 
{\rm min} \left\{
\begin{array}{ll}
1.4 \times 10^{14} M_{\sun} (\frac{x_{\rm i}}{10^{-4}})^{-3/2}
(\frac{L_{\rm LW}}{10^{24} {\rm ergs~s^{-1} Hz^{-1}}})^{3/2}
(\frac{T}{10^3 {\rm K}})^{-6} (\frac{n}{1 {\rm cm^{-3}}})^{-5/4}
(\frac{\Omega_{\rm b}}{0.05})^{-3/4} \\
0.85 \times 10^{11} M_{\sun} 
(\frac{L_{\rm LW}}{10^{24} {\rm ergs~s^{-1} Hz^{-1}}})^{3/2}
(\frac{t_{\rm OB}}{3 \times 10^{6} {\rm yr}})^{3/2}
(\frac{n}{1 {\rm cm^{-3}}}) \\
1.0 \times 10^{8} M_{\sun} (\frac{x_{\rm i}}{10^{-4}})^{-1}
(\frac{L_{\rm LW}}{10^{24} {\rm ergs~s^{-1} Hz^{-1}}})
(\frac{T}{10^3 {\rm K}})^{-1} (\frac{n}{1 {\rm cm^{-3}}})^{-1} 
\end{array}
\right\}
\end{eqnarray}
In equation (\ref{eq:Minf}), each expression corresponds to the case of 
$r^{\rm (inf)}=r_{\rm no-sh}^{\rm (cool)}$, $r_{\rm no-sh}^{\rm (eq)}$,
and $r_{\rm sh}$ from the top to bottom, respectively. 
We shall keep this order hereafter.
At first glance, the first expression in equation (\ref{eq:Minf}) seems 
to be always larger than the others, but its stronger dependence on
temperature makes it important for higher temperature (i.e., more
massive) objects than the normalized value. 
From equation (\ref{eq:Minf}), we can see that the mass within a
region of influence of an O star already exceeds the scale of
the small pregalactic object.   

We have considered in this letter the regulation of star formation by
photodissociation of molecular hydrogen in a pregalactic cloud. 
On the other hand, Lin \& Murray (1992) considered only the regulation by
photoionization.
In such a case, the mass affected by an OB star, namely the baryonic
mass within a Str\"{o}mgren sphere, is
\begin{eqnarray}
M_{\rm b}^{\rm (St)}&=&\frac{m_{\rm p} n Q_{\ast}}{k_{\rm rec} n^{2}}\\
&=& 3.7 \times 10^{3} M_{\sun} (\frac{T}{10^3 {\rm K}})^{0.64} (\frac{n}{1
  {\rm cm^{-3}}})^{-1} (\frac{Q_{\ast}}{10^{49} {\rm s^{-1}}}), 
\end{eqnarray}
where $Q_{\ast}$ is the flux of ionizing photons by a OB star and
$Q_{\ast} \simeq 10^{49} {\rm s^{-1}}$ for an O5 star.
This is by far smaller than our estimated mass of photodissociative
influence $M_{\rm b}^{\rm (inf)}$.
 
To be more specific in the cosmological context, we consider here
pregalactic clouds at virialization.  
The number density at virialization is  
\begin{equation}
\label{eq:nvir}
  n_{\rm vir}=0.68 {\rm cm^{-3}} h_{50}^{2} (\frac{\Omega_{\rm
  b}}{0.05})(\frac{1+z_{\rm vir}}{30})^{3},
\end{equation}
and the virial temperature is 
\begin{equation}
\label{eq:Tvir}
T_{\rm vir}=6.8 \times 10^{2} {\rm K} h_{50}^{2/3} (\frac{\Omega_{\rm
  b}}{0.05})^{-2/3} (\frac{M_{\rm b}}{10^{4}
  M_{\sun}})^{2/3} ( \frac{1+z_{\rm vir}}{30}).
\end{equation}
Substituting equations (\ref{eq:nvir}) and (\ref{eq:Tvir}) into equation
(\ref{eq:Minf}),
we obtain
\begin{equation}
\label{eq:Minf2} 
M_{\rm b}^{\rm (inf)}= 
{\rm min} \left\{
\begin{array}{ll}
2.3 \times 10^{15} M_{\sun} (\frac{x_{\rm i}}{10^{-4}})^{-3/2}
(\frac{L_{\rm LW}}{10^{24} {\rm ergs~s^{-1} Hz^{-1}}})^{3/2} h_{50}^{-13/2}
(\frac{\Omega_{\rm b}}{0.05})^{2} (\frac{M_{\rm b}}{10^{4}
  M_{\sun}})^{-4} ( \frac{1+z_{\rm vir}}{30})^{-39/4} \\
0.58 \times 10^{11} M_{\sun} 
(\frac{L_{\rm LW}}{10^{24} {\rm ergs~s^{-1} Hz^{-1}}})^{3/2}
(\frac{t_{\rm OB}}{3 \times 10^{6} {\rm yr}})^{3/2} h_{50}^{2}
(\frac{\Omega_{\rm b}}{0.05})( \frac{1+z_{\rm vir}}{30})^{3} \\
2.2 \times 10^{8} M_{\sun} (\frac{x_{\rm i}}{10^{-4}})^{-1}
  (\frac{L_{\rm LW}}{10^{24} {\rm ergs~s^{-1} Hz^{-1}}}) h_{50}^{-8/3}
  (\frac{\Omega_{\rm b}}{0.05})^{-1/3} (\frac{M_{\rm b}}{10^{4}
  M_{\sun}})^{-2/3} ( \frac{1+z_{\rm vir}}{30})^{-4} 
\end{array}
\right\}
\end{equation}
In order for the star formation to continue after a massive star forms,
the region of influence must be smaller than the original pregalactic
cloud.
Then $M_{\rm b}^{\rm (inf)}<M_{\rm b}$ is the necessary condition, which
leads to  
\begin{equation}
\label{eq:cond}
M_{\rm b}>
{\rm min} \left\{
\begin{array}{ll}
  1.9 \times 10^{6} M_{\sun}
  (\frac{x_{\rm i}}{10^{-4}})^{-3/10}
  (\frac{L_{\rm LW}}{10^{24} {\rm ergs~s^{-1} Hz^{-1}}})^{3/10}
  h_{50}^{-13/10}
  (\frac{\Omega_{\rm b}}{0.05})^{2/5}
  ( \frac{1+z_{\rm vir}}{30})^{-39/20}\\ 
  0.58 \times 10^{11} M_{\sun} 
  (\frac{L_{\rm LW}}{10^{24} {\rm ergs~s^{-1} Hz^{-1}}})^{3/2}
  (\frac{t_{\rm OB}}{3 \times 10^{6} {\rm yr}})^{3/2} h_{50}^{2}
  (\frac{\Omega_{\rm b}}{0.05})( \frac{1+z_{\rm vir}}{30})^{3}\\
  4.0 \times 10^{6} M_{\sun}
  (\frac{x_{\rm i}}{10^{-4}})^{-3/5}
  (\frac{L_{\rm LW}}{10^{24} {\rm ergs~s^{-1} Hz^{-1}}})^{3/5}
  h_{50}^{-8/5}
  (\frac{\Omega_{\rm b}}{0.05})^{-1/5}
  ( \frac{1+z_{\rm vir}}{30})^{-12/5}
\end{array}
\right\}.
\end{equation}
On the other hand, the baryonic mass of a pregalactic cloud that has virial
temperature $T_{\rm vir}$ is
\begin{equation}
\label{eq:Mb}
M_{\rm b}=1.8 \times 10^{4} M_{\sun} 
(\frac{T_{\rm vir}}{1000 {\rm K}})^{3/2}
h_{50}^{-1} 
(\frac{\Omega_{\rm b}}{0.05})
( \frac{1+z_{\rm vir}}{30})^{-3/2}.
\end{equation}
Comparing equations (\ref{eq:cond}) and (\ref{eq:Mb}), we can see that
for a small  
pregalactic cloud ($T_{\rm vir}<10^{4}$K) the condition $M_{\rm b}^{\rm
 (inf)}<M_{\rm b}$ is hardly satisfied in the redshift range of
$10 \lesssim z_{\rm vir} \lesssim 100$.
This indicates that FUV radiation from one or a few OB stars prohibits the
whole small pregalactic cloud from H$_{2}$ cooling and quenches 
subsequent star formation in it. 

After the death of the first OB star, star formation could occur
somewhere in the cloud, and another OB star could form successively.
Thereafter, some massive stars might form one after another, but only
a few could co-exist simultaneously, as we have shown above.
The timescale for reformation of OB stars depends on that for
${\rm H_2}$ replenishment after the death of the dissociating OB star,
which is 
\begin{eqnarray}
t_{\rm rep}&=&\frac{f^{(\rm cool)}}{k_{\rm H^{-}} x_{\rm e} n} \\
           &=&3 \times 10^{8} {\rm yr} 
           (\frac{n}{1 {\rm cm^{-3}}})^{-3/2} 
           (\frac{T}{10^3 {\rm K}})^{-4} 
           (\frac{x_{\rm e}}{10^{-4}})^{-1} 
           (\frac{\Omega_{\rm b}}{0.05})^{-1/2}.
\end{eqnarray}
If the ionization degree is as high as unity, which is typical in the HII
region, this timescale can be very short, namely,  $t_{\rm rep} \simeq 3
\times 10^{4}$ yr.
If the timescale for reformation of OB stars is smaller than the
lifetime of OB star, it would be possible for a few OB stars to form
every few million years, and, as a result, a considerable amount of
stars would form in a Hubble time.  
However, we do not expect that the star formation continues as long as a
Hubble time, because the gravitational binding energy of
baryonic gas in such a small pregalactic cloud,
\begin{eqnarray}
E_{\rm gr} & \simeq & \frac{G M M_{\rm b}}{R_{\rm vir}} \\
           & \simeq & 3.3 \times 10^{48} {\rm ergs} 
           (\frac{M_{\rm b}}{10^{4} M_{\sun}})^{5/3}
           (\frac{\Omega_{\rm b}}{0.05})^{-2/3} h_{50}^{2/3}
           ( \frac{1+z_{\rm vir}}{30}), 
\end{eqnarray}
where $M$ is the total mass, and $R_{\rm vir}$ is the virial radius of
the cloud, is so small that a few supernova explosions would blow out
such a small pregalactic cloud (see, e.g., Dekel \& Silk 1986). 
Thus, our probable scenario is as follows:
in a small pregalactic cloud,  
star formation occurs only in a photodissociatively regulated fashion
until several supernovae explode, and it stops thereafter.
 
If numerous small stars were formed per a massive star, a substantial
proportion of the original cloud would be converted to stars at last.
However, in the primordial circumstance, the stellar initial mass
function could be strongly biased toward the formation of massive stars
because of the higher temperatures relative to the present-day
counterpart.
If this was the case and OB stars were formed selectively, the amount of 
gas mass that was converted to stars would be extremely small. 
\section{Summary}
We have studied the H$_{2}$ photodissociation region around an OB star in 
a primordial gas cloud.
A region as large as the whole small pregalactic cloud is affected by
only one or a few OB stars and becomes unable to cool in a free-fall time
under the condition appropriate to virialization.
Therefore, in those clouds which have virial temperatures less than
$10^{4}$ K, star formation does not occur efficiently,
unless the primordial initial mass function is extremely weighted
toward low mass stars.

If the reionization of the universe is caused by stellar UV radiation,
some OB stars must form.
However as we have shown, an OB star formed in a small pregalactic cloud
would inevitably photodissociate the whole cloud and subsequent star
formation would be strongly suppressed.
Therefore, stellar UV radiation from small pregalactic clouds cannot
play a significant role in the reionization of the universe.

\acknowledgements
The authors would like to thank Toru Tsuribe for fruitful discussions, 
Humitaka Sato and Naoshi Sugiyama for continuous encouragement, Evan
Scannapieco for checking the English, and the referee, Zolt\'{a}n Haiman, for
a careful reading of the manuscript and for useful comments. 
This work is supported in part by the Grant-in-Aid for Scientific
Research on Priority Areas (No. 10147105) (R.N.), and the Grant-in-Aid for
Scientific Research from the Ministry of Education, Science, Sports and
Culture, No. 08740170 (R.N.).  
%

%%%%%%%%%%%%%%%%%%%%%%%%%%%%%%%%%%%%%%%%%%%%%%%%%%%%%%%%%%%%%%%%%%
%%%%%%  (5) Table           %%%%%%%%%%%%%%%%%%%%%%%%%%%%%%%%%%%%%%
%%%%%%%%%%%%%%%%%%%%%%%%%%%%%%%%%%%%%%%%%%%%%%%%%%%%%%%%%%%%%%%%%%

%%%%%%%%%%%%%%%%%%%%%%%%%%%%%%%%%%%%%%%%%%%%%%%%%%%%%%%%%%%%%%%%%%
%%%%%% (6) References   %%%%%%%%%%%%%%%%%%%%%%%%%%%%%%%%%%%%%%%%%%
%%%%%%%%%%%%%%%%%%%%%%%%%%%%%%%%%%%%%%%%%%%%%%%%%%%%%%%%%%%%%%%%%%

%%%%%%%%%%%%%%%%%%%%%%%%%%%%%%%%%%%%%%%%%%%%%%%%%%%%%%%%%%%%%%%%%%
%%%%%% (7) Figure Captions  %%%%%%%%%%%%%%%%%%%%%%%%%%%%%%%%%%%%%
%%%%%%%%%%%%%%%%%%%%%%%%%%%%%%%%%%%%%%%%%%%%%%%%%%%%%%%%%%%%%%%%%%
%\newpage

%\begin{figure}
%\plotone{f1.eps}
%\caption{
%\label{fig1}}
%\end{figure}

\end{document}